\documentclass[prl,aps,epsfig,superscriptaddress]{revtex4}
\usepackage{amssymb}

\usepackage{bm}
\usepackage{amsfonts}
\usepackage[dvips]{graphicx}
\usepackage{mathrsfs}
\usepackage[intlimits]{amsmath}
\usepackage[colorlinks, citecolor=red]{hyperref}
\usepackage{verbatim}

\newtheorem{theorem}{Theorem}

\begin{document}

\title{Efficient classical simulation of noisy quantum computation}
\author{ Xun Gao}
\affiliation{Center for Quantum Information, IIIS, Tsinghua University, Beijing 100084,
PR China}
\author{Luming Duan}
\affiliation{Center for Quantum Information, IIIS, Tsinghua University, Beijing 100084,
PR China}

\begin{abstract}
Understanding the boundary between classical simulatability and the power of
quantum computation is a fascinating topic. Direct simulation of noisy
quantum computation requires solving an open quantum many-body system, which
is very costly. Here, we develop a tensor network formalism to simulate the
time-dynamics and the Fourier spectrum of noisy quantum circuits. We prove
that under general conditions most of the quantum circuits at any constant
level of noise per gate can be efficiently simulated classically with the
cost increasing only polynomially with the size of the circuits. The result
holds even if we have perfect noiseless quantum gates for some subsets of
operations, such as all the gates in the Clifford group. This surprising
result reveals the subtle relations between classical
simulatability, quantum supremacy, and fault-tolerant quantum computation.
The developed simulation tools may also be useful for solving other open
quantum many-body systems.
\end{abstract}

\maketitle

\section{Introduction}

Any real physical systems are subject to noise due to the inevitable
coupling to the environment. It is important to understand the dynamics of
quantum many-body systems under noise. Universal quantum circuits represent
an important class of quantum many-body systems which are generic in the
sense that they can simulate any other complicated quantum dynamics \cite%
{lloyd1996universal}. Understanding the computational power and the
classical simulatability of noisy universal quantum circuits, therefore,
poses a fascinating question of both fundamental interest and practical
importance \cite{Cirac2012Goals,RevModPhys.86.153,hauke2012can,Harrow2017,1801.00862}.

Direct simulation of noisy quantum many-body dynamics, which is typically
described by an open system through the master equation \cite%
{lindblad1976on,gorini1976completely}, is very challenging, more difficult
than the simulation of the corresponding noiseless dynamics. On the other
hand, it is known that the output of very noisy quantum circuits can be
classically simulated efficiently, i.e., with a polynomial time cost,
through an indirect method \cite{buhrman2006new,kempe2008upper}. If the
noise is not at a very high level, what is the computational power of noisy
quantum devices or systems? Whether is it possible to efficiently simulate
their output or behavior using classical computers? These questions still
remain unknown and they are of great practical importance: with the advent
of the Noisy Intermediate-Scale Quantum (NISQ) era \cite{1801.00862}, most
of the quantum systems that we have in the lab belong to this class where
the noise per quantum gate (operation) is at a modest but non-negligible
level.

In this paper, we address these questions by providing some stimulating
results in this direction. We consider a large class of random universal
quantum circuits under symmetric or asymmetric depolarization noise. For
generic instances from this set of circuits, we prove that if the noise per
quantum gate is at a constant level, the output of these quantum circuits
can be classically efficiently simulated with the time cost polynomial in
the system size. This result holds true even if a particular class of gates
from the universal set are noiseless. For instance, we may assume all the
Clifford gates are noiseless and non-Clifford single-bit gates (such as the $\pi/8$-gate from the universal gate set) have a
constant level of error rate. This assumption is well motivated as the Clifford
gates can be more conveniently protected from noise through fault-tolerant
quantum error correction or topological computation \cite%
{gottesman1997stabilizer,sarma2015majorana,nayak2008non}. For the proof, we
develop a tensor network formalism \cite{Orus2014,biamonte2017tensor} to
analyze the Fourier spectrum of the quantum circuits. The Fourier spectrum
analysis \cite{o2014analysis} has been used recently for classical
simulation of noisy quantum dynamics from the Ising interaction \cite%
{bremner2017achieving,1706.08913}. Our developed tensor network formalism
turns out to be particularly powerful to analyze the Fourier spectrum when
we have a constant level of noise per gate which leads to independent block
summation and efficient truncation in the Fourier space.

This result has some interesting implication on the pursuit of quantum supremacy
demonstration \cite{Harrow2017,Lund2017}. Conventional complexity theory
arguments on quantum supremacy are for noiseless systems in the asymptotic
limit \cite{1203.5813,Lund2017,Harrow2017}, and in this case efficient
classical simulation of certain quantum circuits will lead to collapse of
the so-called polynomial hierarchy in the computational complexity classes
under reasonable conjectures. The real systems will necessarily have noise
and in this case the result here shows that in the asymptotic limit (meaning
that the circuit depth goes to infinity) the output of the circuit is
classically simulatable and actually approaches a uniform distribution for
any constant level of error per gate. Recently, the approach to a uniform
distribution has been supported with numerical evidence \cite%
{boixo2018characterizing,1708.01875}. Here, we give a rigorous analytic
proof of the asymptotic distribution. This result suggests that in the case of noisy gates as for real
experiments, the claim of quantum supremacy is not based on the complexity
theory argument such as the collapse of the polynomial hierarchy, which is
defined in the asymptotic limit, but instead more based on the practical
difficulty to simulate such quantum systems with classical computers for an
intermediate depth of the circuits (at certain intermediate depths of the
circuits it is difficult to simulate the output classically by all the
algorithms known so far, although in the limit of the infinite depth the
output distribution is known and becomes uniform).

\begin{figure}[tbp]
\includegraphics[width=6in]{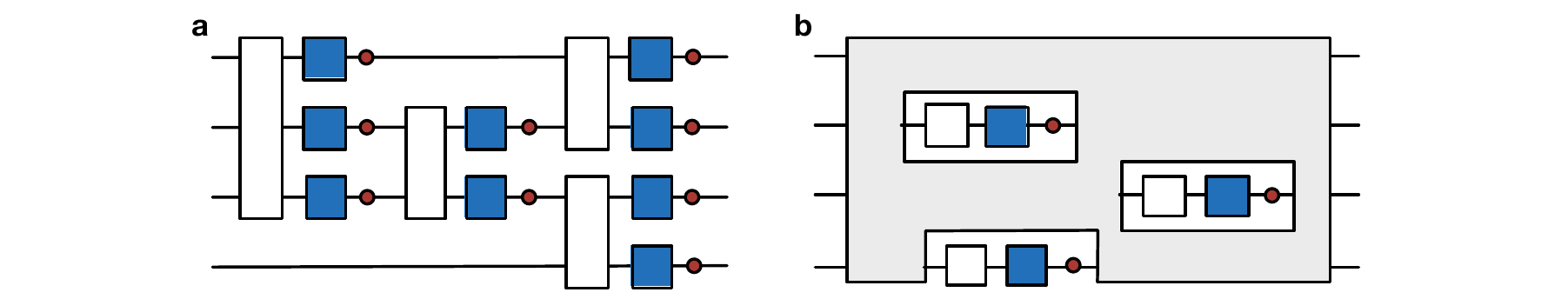}
\caption{ \textbf{Two types of ensembles of noisy quantum circuits.} \textbf{a,} All the gates
represented by blue boxes are chosen uniformly from the four possible Pauli matrices $X^{a}Z^{b}$ with $a,b=0,1$ (including
the identity matrix with $a=b=0$). The gates represented in white could be any gates
in the universal gate set. The red circles represent the channel of noise
that is specified in the main text. The depth $d$ of this circuit is defined as the smallest number
of blue boxes along any horizontal lines. \textbf{b,} The grey color box represents any circuits composed by noiseless Clifford gates.
The white boxes represents non-Clifford gates, which are normally chosen as the single-bit $\pi/8$-gate for universal
quantum computation. The non-Clifford gates are subject to the noise channel represented by the red boxes.}

\label{fig:MT_circuits}
\end{figure}

\section{Summary of the major results}

We model the error for each gate in the quantum circuit by a generalized
type of depolarization noise. For a density operator $\rho $, the noise is
modeled by a composition of Pauli errors $\mathcal{E}_{3}\circ \mathcal{E}%
_{2}\circ \mathcal{E}_{1}(\rho )$, where $\mathcal{E}_{i}(\rho )\equiv
(1-\epsilon _{i})\rho +\epsilon _{i}\sigma _{i}\rho \sigma _{i}$ and {$%
\sigma _{1}=Z$, $\sigma _{2}=X$, $\sigma _{3}=Y$} represent the Pauli
operators along the $z,x,y$ directions. As shown in the Methods section, this error
model allows arbitrary {mixture of Pauli errors including the depolarizing
noise. It is important to have at least two }$\epsilon _{i}$ nonzero for the
noise model to be generic, and we denote the second largest $\epsilon _{i}$ $%
\left( i=1,2,3\right) $ as $\epsilon $, which is a key noise parameter for
our following theorems. The noise channels are denoted as red circles in
Fig. \ref{fig:MT_circuits}.

Instead of analyzing a specific circuit, we consider noisy quantum circuits
randomly chosen from an ensemble and investigate what portions of the
circuits in this ensemble can be efficiently simulated classically. Two
types of circuit ensembles are illustrated in Fig. \ref{fig:MT_circuits}. In
Fig. \ref{fig:MT_circuits}a, we consider an ensemble where all the gates are subject to noise.
To describe this ensemble, for each output qubit of
arbitrary quantum gates, we apply a single-bit operation randomly chosen
from the set of Pauli matrices $\{X^{y_{2}}Z^{y_{1}}\}_{y_{1}y_{2}}$, where
the indices $y_{1},y_{2}$ with values of $0,1$ are uniformly chosen from the
four possible combinations. We use the blue boxes in Fig. \ref%
{fig:MT_circuits}a to represent this random Pauli operation. After the
Pauli operation, each qubit is subject to the noise channel which is denoted
by the red circle in Fig. \ref{fig:MT_circuits}a. A specific quantum circuit
in this ensemble therefore can be parameterized by the sequence of chosen
indices denoted by the vector $\mathbf{y}=\mathbf{y}_{1}\mathbf{y}_{2}$. In Fig. \ref{fig:MT_circuits}b, we consider an ensemble
where part of the gates in the circuit, denoted by the grey block, are
noise-free. The gate set in the grey-block by itself allows an efficient
classical simulation. For instance, we may assume all the Clifford gates are
perfect and thus in the grey block. This assumption is well motivated as the
Clifford gates are easier to be protected from noise by quantum error
correction \cite{gottesman1997stabilizer} or topological computation \cite%
{sarma2015majorana,nayak2008non} and the set of Clifford gates can be
efficiently simulated classically \cite{gottesman1998heisenberg}.
Non-Clifford gates for universal quantum computation, such as the T-gates ($%
\pi /8$-rotations), are subject to noise after the random Pauli operation,
and they are denoted by the white squares in Fig. \ref{fig:MT_circuits}b.

Our major results are summarized by the following two theorems. The first theorem
applies to the circuit ensemble represented by Fig. \ref{fig:MT_circuits}a
where each gate is subject to noise.

\begin{theorem}
\label{thm:random_circuit}\emph{\ Consider the ensemble of noisy quantum
circuits represented by Fig. \ref{fig:MT_circuits}a with circuit depth $d$
and measurement in any local basis on either a constant number of qubits or
all the $n$ qubits with the ensemble satisfying the anti-concentration
condition. {For} at least $1-\eta $ fraction of the circuits in the ensemble
with $\eta =e^{-2\epsilon d}$, the output distribution from the measurement
is approximately a uniform distribution with additive error smaller than $%
\delta =ce^{-\epsilon d}$,{\ where $c$ is a constant independent of }}$n$%
\emph{, }$d$ \emph{and }$\epsilon $\emph{.}
\end{theorem}

In the above theorem, we assume the anti-concentration condition when we
consider the distribution of all the qubits, which is a standard assumption
in the literature on quantum supremacy \cite%
{Lund2017,Harrow2017,aaronson2011computational,PhysRevLett.117.080501,boixo2018characterizing,1706.03786}%
. {Roughly speaking, this condition means that the population is distributed
over the whole Hilbert space of }$2^{n}$-{dimensions and no components
dramatically stand out among the }$2^{n}${\ possibilities. The precise
statement is specified in the Methods section. Theorem 1 means that when all the
gates are subject to noise of a depolarization nature, almost all the
circuits (except for an exponentially small fraction)
will lead to nearly uniform distributions (with exponentially small additive
errors) for the measured qubits when }$\epsilon d$,{\ the product of the
gate error rate and the circuit depth, is large enough. }

Our second theorem applies to the ensemble shown in Fig. \ref%
{fig:MT_circuits}b where part of the gates in the circuit are noise free. In
this case, the output distribution is far from being uniform, however, it
can still be classically simulated within a polynomial time by the following
theorem:

\begin{theorem}
\label{thm:classical_simulation}\emph{\ Consider the ensemble of quantum
circuits shown in Fig. \ref{fig:MT_circuits}b and measurement on a constant
number of qubits. For at least $1-\eta $ fraction of the circuits in this
ensemble, there exists a classical algorithm to approximate the output
probability of the noisy circuits with additive error bounded by $\delta $
which runs in polynomial time
\begin{equation}
\text{poly}(n)\cdot (8m)^{l(\epsilon ,\delta ,\eta )},
\end{equation}%
where $m$ is the number of noisy gates, poly$(n)$ comes from the simulation
cost of noiseless part of the circuit and{\
\begin{equation}
l(\epsilon ,\delta ,\eta )\approx {\ln (2^{r}\delta ^{-1}(1+\sqrt{\eta ^{-1}}%
){)}/2\epsilon ,}
\end{equation}%
where $r$ is the number of qubits being measured and $\epsilon ,\delta ,\eta
$ can be any positive constants.}}
\end{theorem}

Theorem 2 implies that many nontrivial quantum circuits can be efficiently
simulated classically. For instance, even when all the gates are perfect in
the universal gate-set except for the single-bit T gate, which is subject to
a small constant level of error rate, most of universal quantum circuits
composed by these gates are classically simulatable except for
an arbitrarily small constant fraction of subsets. This result is quite
surprising and it has seemingly contradiction with the threshold theorem for
fault-tolerant quantum computation, where we know that the gates below a
small constant level of error rate (threshold) are capable for reliable
universal quantum computing. The result here is actually consistent with the
threshold theorem: it shows generic quantum circuits (except for a \emph{$%
\eta $-}small subset) are classically simulatable under a constant level of
error rate per physical gate. Fault-tolerant quantum error correction requires a
particular circuit structure of physical gates for its implementation, which is
non-generic in the space of all randomly chosen quantum circuits and belongs to the exceptional
small subset that cannot be efficiently simulated classically. If one takes
the logic gates as the basic unit of the circuits, universal quantum
computation means that we can of course realize any generic quantum circuits
of logic qubits. The error rate per logic qubit, however, is not a constant
but required to approach zero inversely with the system size, so again it is outside of the
application region of our theorems here for classical simulatability. Therefore, our result has no
contradiction with the threshold theorem. Instead, it reveals the subtle
boundary between classical simulatability and the power of quantum
computing: although generic noisy quantum circuits are classically
simulatable as indicated by the theorems here, the fault-tolerant quantum
error correction implements a particular (non-generic) type quantum circuit
of physical gates that escapes the curse of noise\ and thus is essential for
the power of quantum computing.

The proof of our theorems has three steps: First, we develop the tensor
network tool to represent the ensemble of noisy quantum circuits and their
Fourier spectrum. Second, we show that the probability distribution from
measurement on the final state can be written as a sum of Fourier series,
where the noise in quantum gates leads to exponential decay of the terms in
this series, so a good approximation can be obtained by truncating the
series. Finally, we show that the tensor network representation of the
Fourier spectrum of quantum circuits breaks into separable pieces, which
makes it possible to efficiently calculate the Fourier series and their
summation. The following three sections are devoted to detailed explanations
of the above three steps one by one. For theorems 1 and 2 with different
circuit ensembles, the proofs only differ in the last step. {For simplicity
of the notation, we focus on the noise model $\mathcal{E}_{2}\circ \mathcal{E}%
_{1}$ for the proof in the main text.} The more general noise model is
discussed in the Methods section.

\section{Tensor network representation of noisy quantum circuits}

\begin{figure}[tbp]
\includegraphics[width=0.8\linewidth]{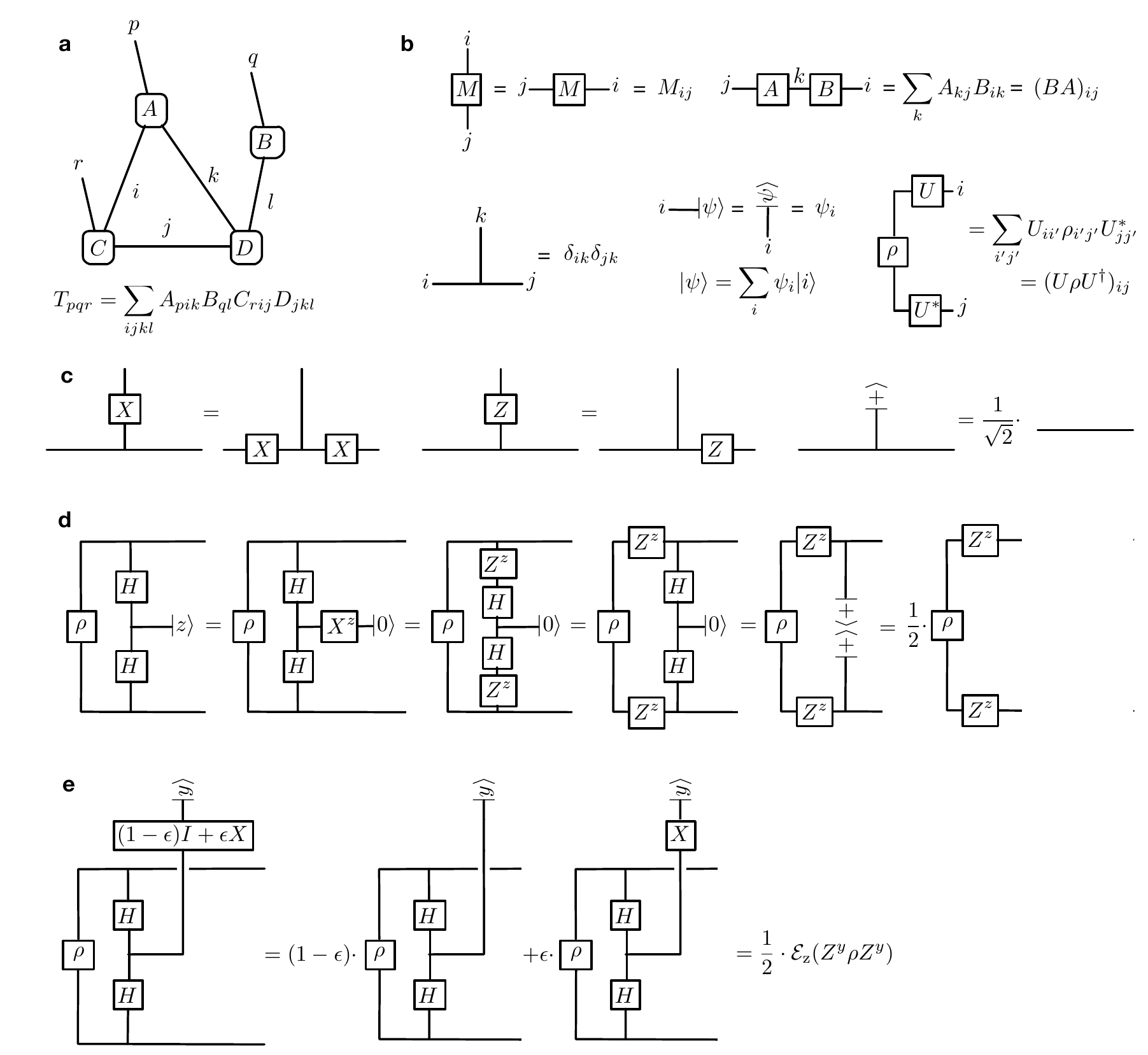}
\caption{ \textbf{Tensor network representation of the noise channel.
a,} Illustration of a tensor network, where the shared edges represent summation
over the corresponding indices ($i,j,k,l$) and the remaining edges represent
free indices ($p,q,r$). \textbf{b,} Basic tensors and notations used in our construction: rank-$2$ tensors $M, A, B$ with
two indices, which are actually matrices (notice that to match the convention of drawing quantum circuits,
the left/right index in the horizontal line corresponds respectively to the right/left index for the matrix;
rank-$3$ identity tensor which constraints all the indices to be equal; rank-$1$ tensors with one index
representing quantum states; a unitary transformation on a density
matrix. \textbf{c,} Basic transformations of a tensor network that are frequently used in our construction, where $X, Z$ are Pauli matrices and $|+\rangle=(|0%
\rangle+|1\rangle)/\protect\sqrt 2$. \textbf{d,} Tensor network
representation of a unitary transformation $Z^{z}$ (with a factor $1/2$, $z=0,1$) on the right side,
which is equivalent to the network on the left side using the above basic transformation rules. \textbf{e,} Tensor
network representation of a quantum noise channel $\mathcal{E}_\mathrm{z}(Z^y%
\protect\rho Z^y)=(1-\protect\epsilon)Z^y\protect\rho Z^y+\protect\epsilon %
Z^{y+1}\protect\rho Z^{y+1}$ (with a factor $1/2$), which is a combination of a unitary
transformation $Z^y$ and a dephasing noise channel $\mathcal{E}_\mathrm{z}$ with the error
probability $\protect\epsilon$ (the phase-flip rate). This representation can
be derived from Fig. \textbf{d} by setting $z=y$ and $z=y+1$, respectively. }
\label{fig:MT_tensors}
\end{figure}

Each particular sample from the quantum circuit ensemble considered in
Theorem 1 or 2 is described by the bit string $\mathbf{y}_{1}\mathbf{y}_{2}$
of $2m$ bits, representing the sequential choice of the Pauli matrices,
where $m$ denotes the total number of noisy gates in the circuit. For this
particular circuit, if we measure $r$ output qubits, the conditional output
distribution can be described as $q_{\mathbf{x}|\mathbf{y}_{1}\mathbf{y}%
_{2}}=q_{\mathbf{x},\mathbf{y}_{1}\mathbf{y}_{2}}/q_{\mathbf{y}_{1}\mathbf{y}%
_{2}}$, where $\mathbf{x}$ denotes the bit string of $r$ measured qubits,
and $q_{\mathbf{x},\mathbf{y}_{1}\mathbf{y}_{2}}$ ($q_{\mathbf{y}_{1}\mathbf{%
y}_{2}}$) denotes respectively the joint output distribution (the marginal
probability to choose this circuit). In our case, the Pauli matrix indices $%
\mathbf{y}_{1}\mathbf{y}_{2}$ are chosen uniformly for the ensemble, so $q_{%
\mathbf{y}_{1}\mathbf{y}_{2}}=2^{-2m}$. If all the gates are perfect (in the
case of $\epsilon=0$), the ideal unitary transformation represented by the
whole quantum circuit is denoted by $U_{\mathbf{y}_{1}\mathbf{y}_{2}}$. With
noise for the gates, the corresponding transformation is denoted by the
quantum channel super-operator $\Phi _{\mathbf{y}_{1}\mathbf{y}_{2}}$. We
then define two conditional distributions with
\begin{equation}
q_{\mathbf{x}|\mathbf{y}_{1}\mathbf{y}_{2}}\equiv \left\vert (\langle
\mathbf{x}|\otimes I)U_{\mathbf{y}_{1}\mathbf{y}_{2}}|0\rangle ^{\otimes
n}\right\vert ^{2}  \label{1}
\end{equation}%
and
\begin{equation}
q_{\mathbf{x}|\mathbf{y}_{1}\mathbf{y}_{2}}^{\prime }\equiv \mbox{tr}\left[
\Phi _{\mathbf{y}_{1}\mathbf{y}_{2}}\left( (|0\rangle \langle 0|)^{\otimes
n}\right) (|\mathbf{x}\rangle \langle \mathbf{x}|\otimes I)\right]
\end{equation}%
where $|0\rangle ^{\otimes n}$ denotes the initial state of $n$ qubits.

Tensor network provides a powerful diagrammatic method to represent quantum
states and their evolution under noisy quantum circuits. For the
distribution $q_{\mathbf{x},\mathbf{y}_{1}\mathbf{y}_{2}}$, it
only involves unitary transformations. In Fig. 2, we show some basic notations
of tensor network units and their composition rules. As a noiseless quantum
circuit is composed of elementary unitary gates, it is straightforward to
represent the distribution $q_{\mathbf{x},\mathbf{y}_{1}\mathbf{y}_{2}}$
with a tensor network. Different from $q_{\mathbf{x},\mathbf{y}_{1}\mathbf{y}%
_{2}}$, to represent $q_{\mathbf{x},\mathbf{y}_{1}\mathbf{y}_{2}}^{\prime }$%
, we also need to have a tensor network representation of the dephasing and
the bit flip error channels. Using the tensor network diagrammatic method, we
derive in Fig. 2d and 2e the representation for the dephasing error channel $%
\mathcal{E}_{1}(Z^{y}\rho Z^{y})$ ($y=0,1$). Similarly, with additional
Hadamard gates, we have a tensor network representation for the bit flip
error channel. The noise in our circuit can be composed from these basic
units. In Fig. 3, we show the tensor network representation of the random
gate set $X^{y_{2}}Z^{y_{1}}$ (Fig. 3b) as well as this gate set followed by
a noise channel $\mathcal{E}_{2}\circ \mathcal{E}_{1}$ (Fig. 3a). By the
composition of these units, we get the representation for the probability
distribution $q_{\mathbf{x}|\mathbf{y}_{1}\mathbf{y}_{2}}^{\prime }$.

\begin{figure}[tbp]
\includegraphics[width=0.8\linewidth]{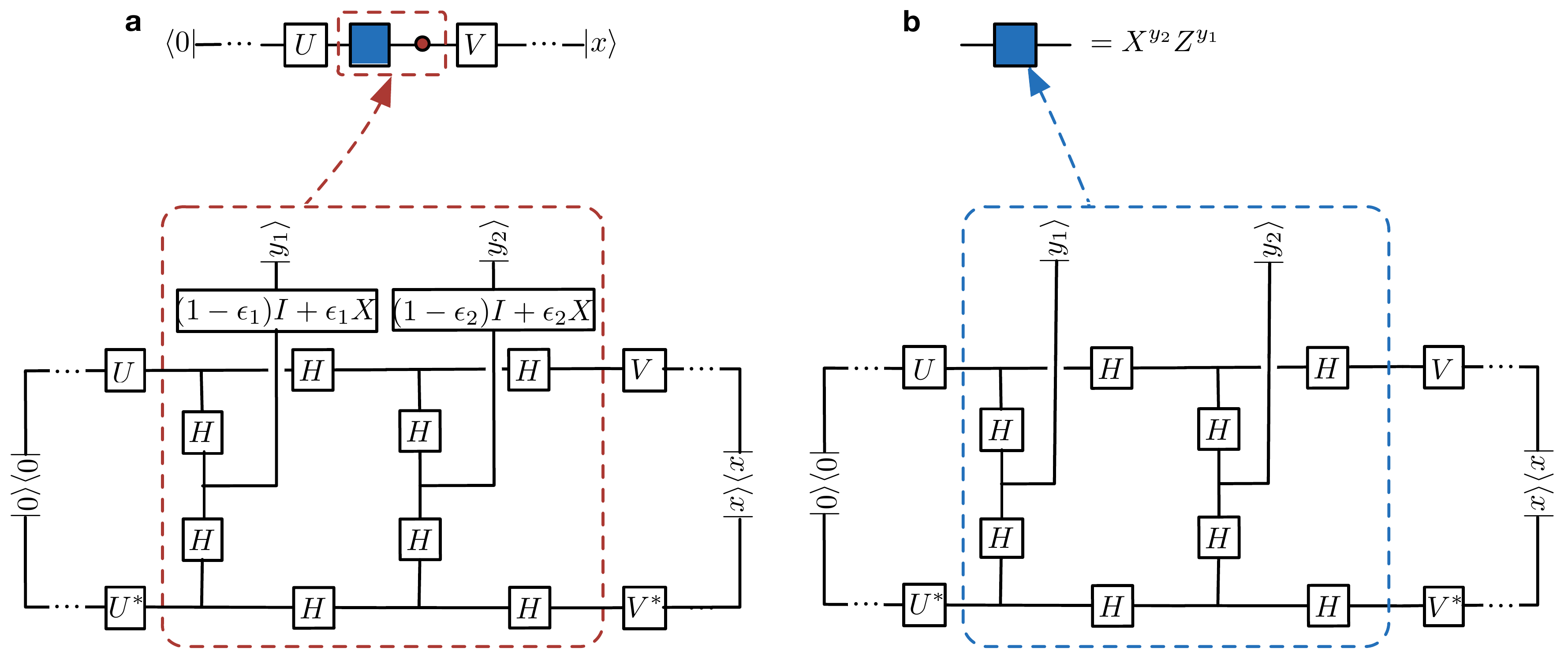}
\caption{ \textbf{Tensor network representation of the basic units of
noisy quantum circuits.}
\textbf{a,} Representation of the Pauli matrix ensemble (the blue box) followed by the noise channel (the red dot). The blue box
denotes the ensemble of gates $\{X^{y_2}Z^{y_1}\}$ with $y_1,y_2$ chosen randomly from $\{0,1\}$ with equal probability.
The red dot denotes the noise channel $\mathcal{E}_\mathrm{x}(\mathcal{E}_\mathrm{z}(\protect\rho%
))$, a combination of phase flip and bit-flip errors. The tensor network representation of this noise channel is derived from Fig. 2e by noticing
that the additional bit flip channel is obtained through the Hadamard transformation $H$ with
$HZH=X$ and $H\mathcal{E}_\mathrm{z}(Z^y%
\protect\rho Z^y)H=\mathcal{E}_\mathrm{x}(X^y%
\protect H\rho H X^y)$. We assume the noisy quantum circuit starts at an initial state denoted as $|0\rangle$ and ends with a measurement
in the computational basis $|x\rangle$, and the contraction
of the tensor network gives the joint probability distribution $q^\prime_{x,y_1,y_2}$.
\textbf{b,} Representation of the noiseless Pauli matrix ensemble (the blue box only). Contraction of the tensor
network in this case gives the joint probability distribution $q_{x,y_1,y_2}$ for the corresponding noiseless circuit.
}
\label{fig:MT_ensemble}
\end{figure}

\section{Fourier analysis of noisy quantum circuits and its tensor network
representation}

\begin{figure}[tbp]
\includegraphics[width=0.8\linewidth]{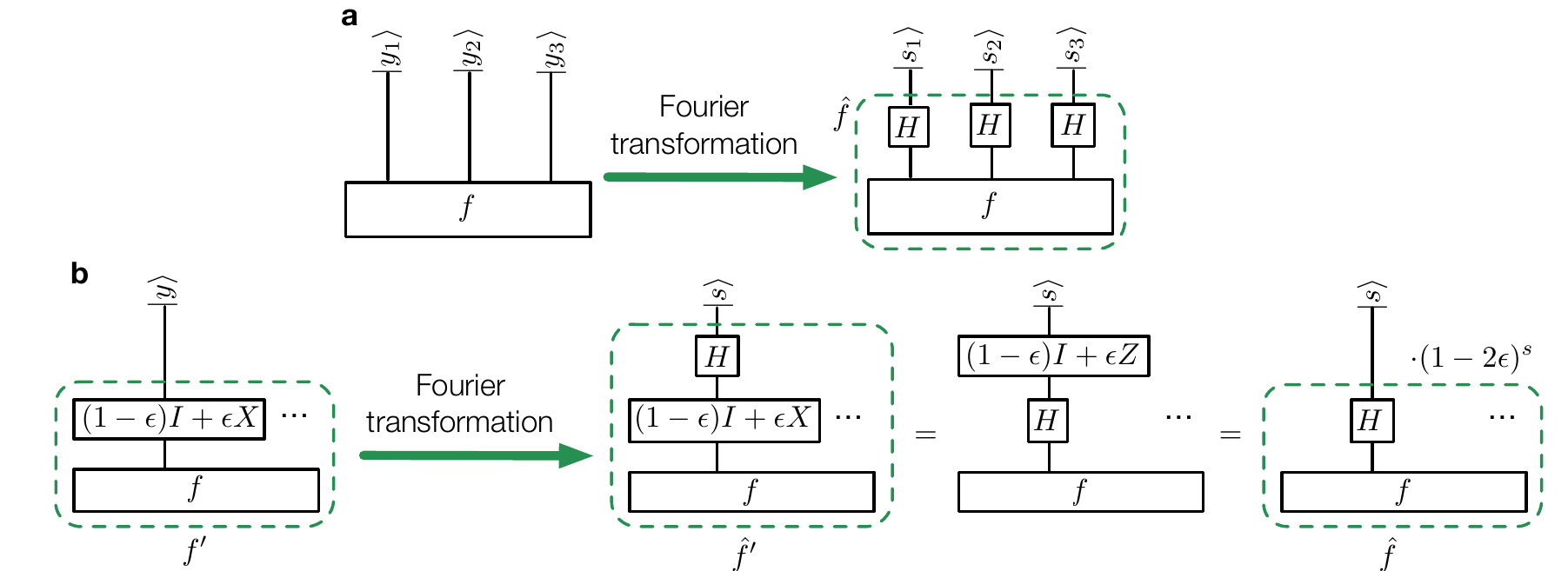}
\caption{ \textbf{Tensor network representation of Fourier transformation of noisy quantum circuits. a,}
Tensor Representation of Fourier transformation on boolean variables $y_i$, which is equivalent to the Hadamard
transformation on the corresponding indices. The inverse Fourier transformation is the same because of $H^{2}=I$.
\textbf{b, } Representation of the Fourier transformation on noisy circuits. As $HXH=Z$ and $[(1-\protect\epsilon )I+\protect%
\epsilon Z]|s\rangle =(1-2\protect\epsilon )^{s}|s\rangle $, we immediately get that the Fourier spectrum of the noisy circuit is
related to the spectrum of the noiseless circuit by a simple multiplication factor $(1-2\protect\epsilon )^{s}$. }
\label{fig:MT_FT}
\end{figure}

Each sample from the ensemble of noisy quantum circuits depends on a
sequence of binary variables $\mathbf{y\equiv y}_{1}\mathbf{y}_{2}$ that
specify its choice of Pauli gates. For the probability distributions $q_{%
\mathbf{x},\mathbf{y}_{1}\mathbf{y}_{2}}$ and $q_{\mathbf{x},\mathbf{y}_{1}%
\mathbf{y}_{2}}^{\prime }$, one can define a binary Fourier transformation
with respect to their $\mathbf{y}$ variables as follows \cite%
{o2014analysis,bremner2017achieving}:
\begin{eqnarray}
\hat{q}_{\mathbf{x},\mathbf{s}_{1}\mathbf{s}_{2}} &=&\frac{1}{2^{m}}\sum_{%
\mathbf{y}}(-1)^{\mathbf{s}\cdot \mathbf{y}}q_{\mathbf{x,y}_{1}\mathbf{y}%
_{2}},\quad  \notag \\
q_{\mathbf{x,y}_{1}\mathbf{y}_{2}} &=&\frac{1}{2^{m}}\sum_{\mathbf{s}}(-1)^{%
\mathbf{s}\cdot \mathbf{y}}\hat{q}_{\mathbf{x},\mathbf{s}_{1}\mathbf{s}_{2}},
\end{eqnarray}%
where $\mathbf{s}=\mathbf{s}_{1}\mathbf{s}_{2}$ of $2m$ bits are the dual
variables of $\mathbf{y}=\mathbf{y}_{1}\mathbf{y}_{2}$ and $\mathbf{s}\cdot
\mathbf{y}=\mathbf{s}_{1}\cdot \mathbf{y}_{1}+\mathbf{s}_{2}\cdot \mathbf{y}%
_{2}$ denotes the binary inner product. The Fourier transformation $\hat{q}_{%
\mathbf{x},\mathbf{s}_{1}\mathbf{s}_{2}}^{\prime }$ is defined in the same
way. An important observation here is that the binary Fourier transformation
has a very natural tensor network representation as shown in Fig. \ref%
{fig:MT_FT}a. By identifying the variables of a boolean function as tensor
indices, the binary Fourier transformation is simply contraction with the
Hadamard matrices, which is also known as the Walsh-Hadamard transformation.

Using the tensor network representation of the binary Fourier
transformation, it is easy to see a relation between $\hat{q}_{\mathbf{x},%
\mathbf{s}_{1}\mathbf{s}_{2}}^{\prime }$ and $\hat{q}_{\mathbf{x},\mathbf{s}%
_{1}\mathbf{s}_{2}}$ by contraction with the noise matrix $(1-\epsilon
_{1,2})I+\epsilon _{1,2}X$ as shown in Fig. \ref{fig:MT_FT}b. We get
an important property \cite{o2014analysis,bremner2017achieving}
\begin{eqnarray}
\hat{q}_{\mathbf{x},\mathbf{s}_{1}\mathbf{s}_{2}}^{\prime } &=&(1-2\epsilon
_{1})^{|\mathbf{s}_{1}|}(1-2\epsilon _{2})^{|\mathbf{s}_{2}|}\hat{q}_{%
\mathbf{x},\mathbf{s}_{1}\mathbf{s}_{2}}  \notag \\
&\simeq &e^{-2(\epsilon _{1}|\mathbf{s}_{1}|+\epsilon _{2}|\mathbf{s}_{2}|)}%
\hat{q}_{\mathbf{x},\mathbf{s}_{1}\mathbf{s}_{2}}
\end{eqnarray}%
where $|\mathbf{s}_{1,2}|$ denotes the Hamming weight of the bit string $%
\mathbf{s}_{1,2}$, i.e., the number of $1$s in $\mathbf{s}_{1,2}$. So the
noise in gates makes the Fourier components exponentially decay with $|%
\mathbf{s}_{1,2}|$, which suggests that the summation can be approximated by
truncating the Fourier series under noisy gates.

We define the pseudo probability $p_{\mathbf{x},\mathbf{y}_{1}\mathbf{y}%
_{2}}^{\prime }$ through truncating the Fourier series of $\hat{q}_{\mathbf{x%
},\mathbf{y}_{1}\mathbf{y}_{2}}^{\prime }$ to the Hamming weight $l$ with $%
\hat{p}_{\mathbf{x},\mathbf{s}_{1}\mathbf{s}_{2}}^{\prime }=\hat{q}_{\mathbf{%
x},\mathbf{s}_{1}\mathbf{s}_{2}}^{\prime }$ if $|\mathbf{s}_{1}|+|\mathbf{s}%
_{2}|<l$ and $\hat{p}_{\mathbf{x},\mathbf{s}_{1}\mathbf{s}_{2}}^{\prime }=0$
otherwise. Assuming each $\hat{p}_{\mathbf{x},\mathbf{s}_{1}\mathbf{s}%
_{2}}^{\prime }$ can be efficiently computed classically within a time $t$
(we will prove this in the next section), then computing $p_{\mathbf{x},%
\mathbf{y}_{1}\mathbf{y}_{2}}^{\prime }$ through the Fourier inverse
transformation is $t(2m)^{l}$ since there are at most $(2m)^{l}$ non-zero
terms in the summation. The quantity we need is $q_{\mathbf{x}|\mathbf{y}_{1}%
\mathbf{y}_{2}}^{\prime }=4^{m}q_{\mathbf{x},\mathbf{y}_{1}\mathbf{y}%
_{2}}^{\prime }$, which is approximated by $p_{\mathbf{x}|\mathbf{y}_{1}%
\mathbf{y}_{2}}^{\prime }=4^{m}p_{\mathbf{x},\mathbf{y}_{1}\mathbf{y}%
_{2}}^{\prime }$. To quantify this approximation, we use the additive error
defined as $\delta _{\mathbf{y}_{1}\mathbf{y}_{2}}\equiv \sum_{\mathbf{x}%
}|p_{\mathbf{x}|\mathbf{y}_{1}\mathbf{y}_{2}}^{\prime }-q_{\mathbf{x}|%
\mathbf{y}_{1}\mathbf{y}_{2}}^{\prime }|.$ Instead of analyzing each
individual $\delta _{\mathbf{y}_{1}\mathbf{y}_{2}}$, we analyze the average
behavior through the expectation value $\delta _{0}\equiv \mathbb{E}_{%
\mathbf{y}_{1}\mathbf{y}_{2}}[\delta _{\mathbf{y}_{1}\mathbf{y}_{2}}]$ and
the variance $\Delta \equiv \sqrt{\mathbb{E}_{\mathbf{y}_{1}\mathbf{y}%
_{2}}[\delta _{\mathbf{y}_{1}\mathbf{y}_{2}}^{2}]-\delta _{0}^{2}}$, where $%
\mathbf{y}_{1}\mathbf{y}_{2}$ are chosen from the uniform distribution. By
direct calculation shown in the Methods section, we find that
\begin{equation}
\delta _{0}\leq ce^{-2\epsilon l},\quad \Delta \leq ce^{-2\epsilon l},
\label{eq:DE}
\end{equation}%
where $\epsilon =\min (\epsilon _{1},\epsilon _{2})$ and $c$ is a constant.
By using the Chebyshev inequality, we can then bound $\delta _{\mathbf{y}_{1}%
\mathbf{y}_{2}}\leq \delta $ for at least $1-\eta $ fraction of $\mathbf{y}%
_{1}\mathbf{y}_{2}$ as long as $\eta (\delta -\delta _{0})^{2}\leq \Delta
^{2}$. By choosing arbitrarily small $\eta $ and $\delta $, we conclude that
almost all $\delta _{\mathbf{y}_{1}\mathbf{y}_{2}}$ are well bounded and the
approximation is valid for most of the random quantum circuits.

\begin{figure}[tbp]
\includegraphics[width=0.8\linewidth]{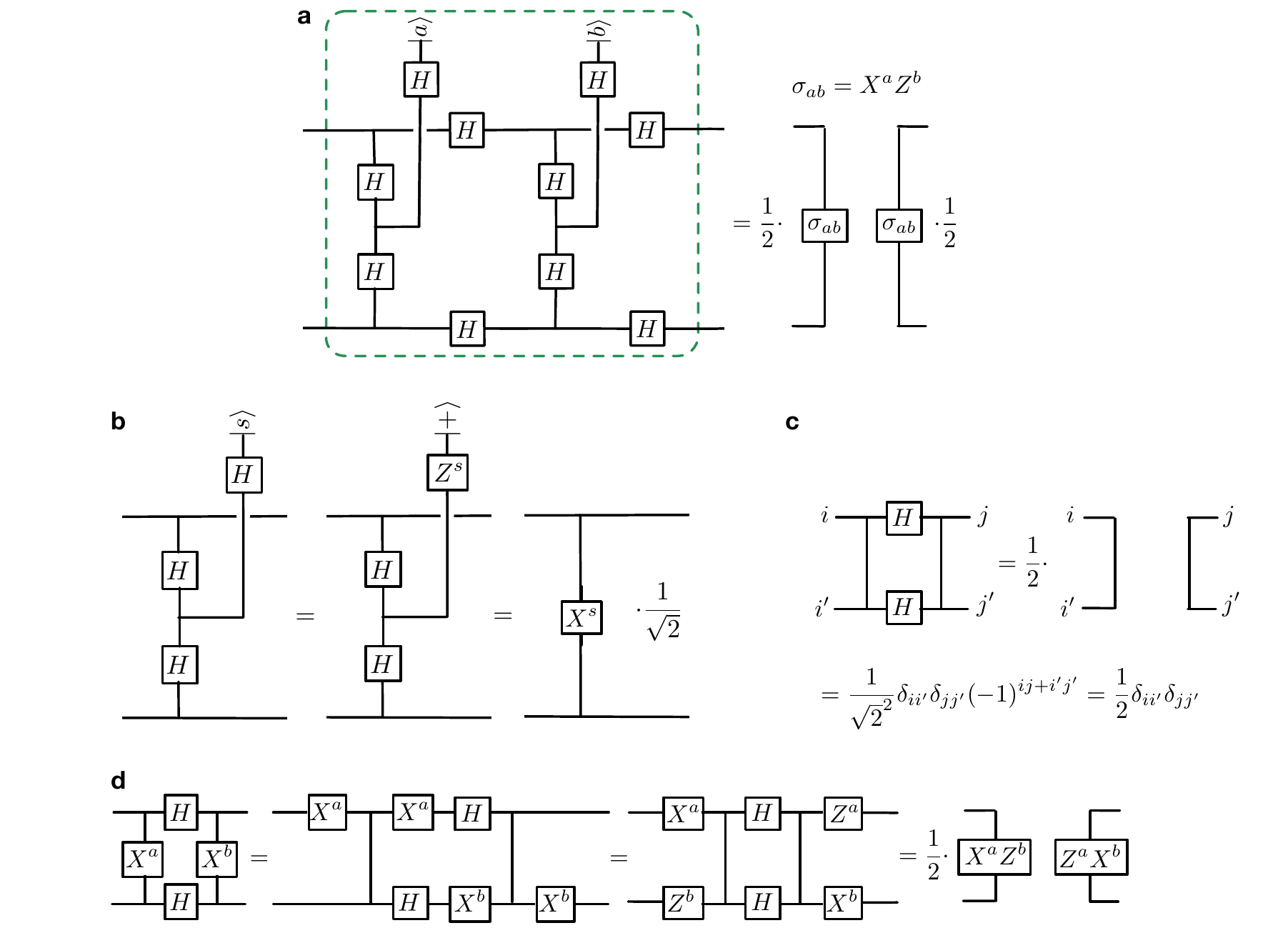}
\caption{ \textbf{Separability for the Tensor network representation of the Fourier spectrum of quantum circuits.}
\textbf{a,} The pivotal property of separability for the Fourier spectrum, which shows that the Fourier transformation of the Pauli ensemble (the basic unit denoted by the blue box in Fig. 3b) on the left side can be written as a product of two separable tensor
networks on the right side, where $\sigma_{ab}$ denotes the matrix $X^{a}Z^{b}$. This separability is
derived in Figs. \textbf{b-d} through the diagrammatic method.
The separability is the key to simplification of the calculation of the Fourier spectrum
for ensembles of large quantum circuits. \textbf{b,} Step 1 to derive the separability which uses the
equalities in Fig. \protect\ref{fig:MT_tensors}c. \textbf{c,} Step 2 to
derive the separability in the case with $a=b=0$. \textbf{d,} Step 3 to derive
the separability for the general case with arbitrary $a,b$. We have used the result of Fig. 5b  and the tensor property in Fig. 2c.}
\label{fig:MT_pivotal}
\end{figure}

\section{Proof of theorems through analyzing the Fourier spectrum}

\begin{figure}[tbp]
\includegraphics[width=0.8\linewidth]{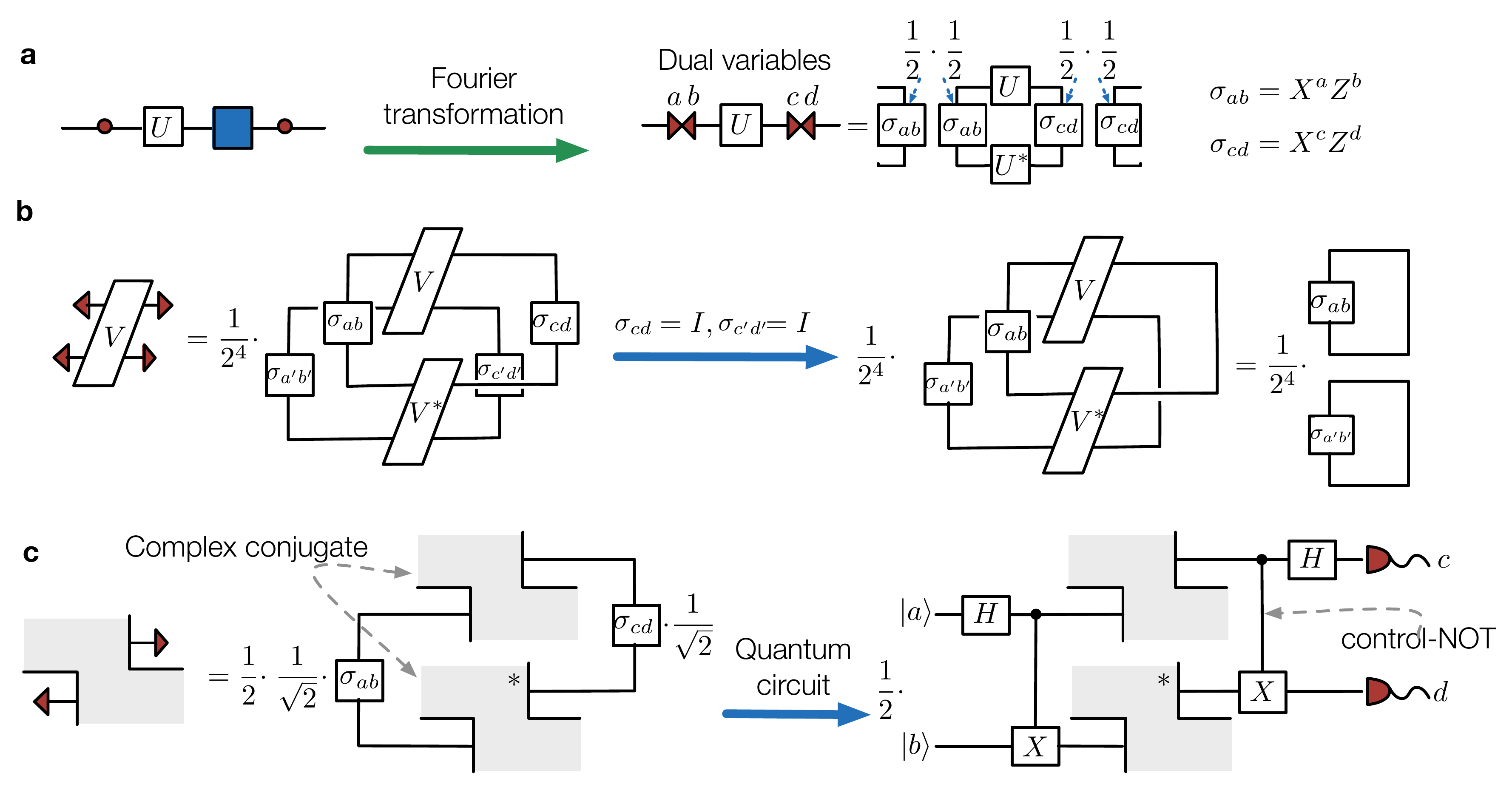}
\caption{ \textbf{The computing rule for the Fourier spectrum through the tensor network representation.} \textbf{a,}
The computing rule for the Fourier spectrum of noisy quantum circuits: after the transformation, the tensor network breaks into
two separated pieces at the position of the basic unit of Paul ensemble (the blue box and the red dot), so we
denote the breaking point by the two red triangles, which correspond to a product of the tensor networks shown on the right side.
\textbf{b,} An important property for contraction of the tensor network used for proof of Theorem 1. When we contract a piece of tensor network with an arbitrary two-qubit gate $V$ in the between, if $%
\protect\sigma_{cd}=\protect\sigma_{c^\prime d^\prime}=I$ on one side of the network, the
contraction leads to a tensor network shown on the right side which is proportional to $\mbox{tr}\protect\sigma_{ab} \cdot\mbox{tr}\protect\sigma
_{a^\prime b^\prime}$ and non-zero only when $\protect\sigma_{ab}=\protect%
\sigma_{a^\prime b^\prime}=I$. \textbf{c,} A property of the tensor network used for proof of Theorem 2.  The grey part represents any Clifford quantum circuits and the lower grey part denotes the complex conjugate of the upper one. The matrices $\sigma_{ab}$ and $\sigma_{cd}$ can be
represented as the Clifford gates (the Hadamard H and the CNOT gate) acting on the computational basis-vectors $|a\rangle, |b\rangle$ (on the left) or measured in the computational basis $|c\rangle, |d\rangle$ (on the right). The whole equivalent circuit shown on the right figure is then of Clifford type and can be efficiently simulated classically.}
\label{fig:MT_rules}
\end{figure}

Through the above analysis, we reduce the problem of calculating the
conditional distribution $q_{\mathbf{x}|\mathbf{y}_{1}\mathbf{y}%
_{2}}^{\prime }$ to computing of the Fourier spectrum $\hat{q}_{\mathbf{x},%
\mathbf{s}_{1}\mathbf{s}_{2}}$. To prove theorems 1 and 2, the remaining
task is to find a method to efficiently compute $\hat{q}_{\mathbf{x},\mathbf{%
s}_{1}\mathbf{s}_{2}}$. In Fig. \ref{fig:MT_pivotal}, using the diagrammatic
method, we derive the tensor network representation of the Fourier spectrum $%
\hat{q}_{\mathbf{x},\mathbf{s}_{1}\mathbf{s}_{2}}$. The final representation
is shown in Fig. 5a, with the derivation steps illustrated in Fig. 5(b-d).
The tensor network representation of $\hat{q}_{\mathbf{x},\mathbf{s}_{1}%
\mathbf{s}_{2}}$ shows a crucial property: it breaks into separable pieces
at every places of the random Pauli gates. This property allows us to
efficiently calculate the Fourier spectrum $\hat{q}_{\mathbf{x},\mathbf{s}%
_{1}\mathbf{s}_{2}}$ for the whole circuit.\

First, we prove Theorem 1 by calculating the Fourier spectrum $\hat{q}_{%
\mathbf{x},\mathbf{s}_{1}\mathbf{s}_{2}}$ for the ensemble of the noisy
circuits shown in Fig. \ref{fig:MT_circuits}a. The Fourier component $\hat{q}%
_{\mathbf{x},\mathbf{s}_{1}\mathbf{s}_{2}}$ breaks into product of many
independent fragments, separated by the blue boxes shown in Fig. 6a (see
also Figs. 1a, 3a, and 5a for the representation). This product is nonzero
if and only if all the fragments are nonzero. The fragment of a single-bit
gate is shown in Fig. 6a. To make this fragment nonzero, if $\sigma _{ab}=I$
(i.e., the dual variables $a=b=0$ in the Fourier index $\mathbf{s}_{1}%
\mathbf{s}_{2}$), we should also have $\sigma _{cd}=I$. Similarly, as shown
in Fig. 6b, to make the Fourier fragment of an entangling gate nonzero, if $%
\sigma _{ab}=\sigma _{a^{\prime }b^{\prime }}=I$, we have to choose $\sigma
_{cd}=\sigma _{c^{\prime }d^{\prime }}=I$. Using this property, we can prove
that the Fourier spectrum $\hat{q}_{\mathbf{x},\mathbf{s}_{1}\mathbf{s}%
_{2}}=0$ for all the components $\mathbf{s}_{1}\mathbf{s}_{2}$ with $0<|%
\mathbf{s}_{1}|+|\mathbf{s}_{2}|<d$. To prove this, we cut the circuit in
Fig. 1a into $d$ vertical layers, where $d$ is the circuit depth. If we have
the Fourier index $|\mathbf{s}_{1i}|+|\mathbf{s}_{2i}|$ $=0$\quad for any $i$%
th layer (corresponding to choice of the $I$ matrix for all the cuts in this
layer), with the property shown in Fig. 6a and 6b, we have to choose $|%
\mathbf{s}_{1j}|+|\mathbf{s}_{2j}|$ $=0$ for all the other layers $j$ to
make the product $\hat{q}_{\mathbf{x},\mathbf{s}_{1}\mathbf{s}_{2}}$ nonzero
(otherwise one of its Fourier fragments will be zero). So, to make $\hat{q}_{%
\mathbf{x},\mathbf{s}_{1}\mathbf{s}_{2}}$ nonzero, we either have the
Hamming weight $|\mathbf{s}_{1}|+|\mathbf{s}_{2}|=0$, or have $|\mathbf{s}%
_{1i}|+|\mathbf{s}_{2i}|$ $>0$ for each of the $d$ layers which means $|%
\mathbf{s}_{1}|+|\mathbf{s}_{2}|\geq d$. This proves the above statement.

To finish the proof of theorem 1, we just need to take $l=d$ in Eq. (\ref%
{eq:DE}). This leads to $\hat{p}_{\mathbf{x},\mathbf{s}_{1}\mathbf{s}%
_{2}}^{\prime }=0$ for $|\mathbf{s}_{1}|+|\mathbf{s}_{2}|>0$, which gives a
uniform distribution $p_{\mathbf{x}|\mathbf{y}_{1}\mathbf{y}_{2}}^{\prime }$
after the Fourier transformation. In combination with the Chebyshev
inequality, we thus have proved theorem 1 by choosing $\delta \sim e^{-\epsilon
d}\gg \delta _{0}\sim e^{-2\epsilon d}$.

Now we prove theorem 2 by considering the ensemble of noisy quantum circuits
shown in Fig. \ref{fig:MT_circuits}b. The Fourier spectrum of this circuit
is illustrated in Fig. \ref{fig:MT_rules}c, where the grey part denotes the
Clifford gates. The Fourier circuit breaks up at each place of the
non-Clifford gates (the white and blue boxes in Fig. 2a), but due to the
connection by the Clifford gates, it does not break into independent pieces.
In Fig. 6c, we write the matrix at the breaking points into an equivalent
Clifford gates. So the whole circuit for the Fourier spectrum becomes almost
Clifford except for the non-Clifford gates (white boxes) at the breaking
points. We can decompose the non-Clifford gates into a summation of $4^{l}$
terms where each term is a tensor product of Pauli matrices. Only $l$ white
boxes have non-trivial effect because the non-Clifford gate $U$ in Fig. \ref%
{fig:MT_rules}c cancels through $U^{\dag }IU=I$ if the corresponding dual
variables in $\mathbf{s}_{1}\mathbf{s}_{2}$ are both $0$ (i.e., $a=b=0$ in
Fig. 6c). The Fourier spectrum is therefore expressed as a summation of a
constant number ($4^{l}$) of Clifford circuits, and by the Gottesman-Knill
theorem \cite{gottesman1998heisenberg}, each Clifford circuits can be
efficiently simulated.

{To count the time complexity of computing $\hat{p}_{\mathbf{x}|\mathbf{y}%
_{1}\mathbf{y}_{2}}^{\prime }$, suppose we truncate the Fourier series to
the Hamming weight $l$. The time complexity of computing each Fourier
component is $4^{l}t$, where $4^{l}$ originates from the Pauli decomposition of
the non-Clifford gates as we have discussed above and $t=\text{poly}(n)$ is
the time complexity of computing the remaining Clifford circuits. Then there
are at most $(2m)^{l}$ terms to be summarized, so the time complexity is $%
\text{poly}(n)(8m)^{l}$. Given the required precision of the output $%
1-\delta $ and the probability of failure $\eta $, we have  $\delta \precsim
\Delta /\sqrt{\eta }+\delta _{0}\precsim (1+1/\sqrt{\eta })ce^{-2\epsilon l}$
according to the Chebyshev inequality, where $c=\sqrt{2^{r}}$ is a constant
and $r$ is the number of qubits being measured. This requires $l\approx \ln
(c\delta ^{-1}(1+\sqrt{\eta ^{-1}}))/2\epsilon $, which gives the expression
in theorem 2. This completes the proof of the theorem 2.

\section{Summary and Outlook}

In summary, we have proved that most of noisy quantum circuits can be
efficiently simulated classically under any constant level of error rate per
gate. The result holds even if a subset of quantum gates, such as all
the Clifford gates, can be done perfectly. The seemingly contradiction
between this surprising result and the threshold theorem for fault
tolerant quantum computation is resolved by noting that the quantum error
correction circuit is required to be highly structured and thus may take
only a zero-measure subspace under random choice of quantum circuits. The theorems derived in this paper show that 
for noisy quantum circuits the dividing lines between classical simulatability, quantum
supremacy, and universal quantum computing
are very subtle. What are the characteristic features of noisy quantum circuits to make
them either classically simulatable or noise-resilient for universal quantum computing? It remains an important open question 
and further studies along this line will surely
deepen our understanding of the power and limitation of
quantum computation.

To prove our theorems, we have developed a powerful tool based on
tensor network representation of noisy quantum circuits and their
measurement outcomes. We discover that the tensor network representation
of the Fourier spectrum of quantum circuits always breaks into many
independent pieces. This separability of its Fourier spectrum is a
surprising and nice property, which greatly simplifies our analyses of
complicated dynamics of quantum circuits. Besides its use in analyzing the structure of
large-scale quantum circuits, this property of separable Fourier spectrum
may find applications in other fields, such as for understanding of many-body quantum dynamics under
open environments \cite{Cirac2012Goals,RevModPhys.86.153,hauke2012can}, a key
problem on the physics frontier.

\section{Methods}

\subsubsection{Noise model}

We can simulate the more general noise model from an arbitrary mixture of
Pauli errors
\begin{equation}
\mathcal{E}(\rho )=(1-\epsilon _{\mathrm{x}}-\epsilon _{\mathrm{y}}-\epsilon
_{\mathrm{z}})\rho +\epsilon _{\mathrm{x}}X\rho X+\epsilon _{\mathrm{y}%
}Y\rho Y+\epsilon _{\mathrm{z}}Z\rho Z
\end{equation}%
by successively applying the noise channel $\mathcal{E}_{\mathrm{z}}$
\begin{equation}
(S\cdot H\cdot \mathcal{E}_{\mathrm{z}}^{(3)}\cdot H\cdot S^{\dag })\circ
(H\cdot \mathcal{E}_{\mathrm{z}}^{(2)}\cdot H)\circ \mathcal{E}_{\mathrm{z}%
}^{(1)},
\end{equation}%
with
\begin{equation}
\mathcal{E}_{\mathrm{z}}^{(i)}(\rho )=(1-\epsilon _{i})\rho +\epsilon
_{i}Z\rho Z,
\end{equation}%
where $\circ $ denotes composition of quantum channels, $\cdot $ denotes
matrix multiplication, and $H$ ($S$) denotes respectively the Hadamard
(phase) gate. Since $HZH=X$ and $SHZHS^{\dag }=Y$, adjusting $\epsilon _{i}$
will give any non-zero small $\epsilon _{\mathrm{x},\mathrm{y},\mathrm{z}}$
through the following relations
\begin{eqnarray}
\epsilon _{\mathrm{z}} &=&\epsilon _{1}(1-\epsilon _{2})(1-\epsilon
_{3})+(1-\epsilon _{1})\epsilon _{2}\epsilon _{3};  \notag \\
\epsilon _{\mathrm{x}} &=&\epsilon _{2}(1-\epsilon _{1})(1-\epsilon
_{3})+(1-\epsilon _{2})\epsilon _{1}\epsilon _{3};  \notag \\
\epsilon _{\mathrm{y}} &=&\epsilon _{3}(1-\epsilon _{1})(1-\epsilon
_{2})+(1-\epsilon _{3})\epsilon _{1}\epsilon _{2}.
\end{eqnarray}%
The value of the Jacobian determinant for the above transformation is
\begin{equation}
(2\epsilon _{1}-1)(2\epsilon _{2}-1)(2\epsilon _{3}-1).
\end{equation}%
According to the inverse function theorem, the solution always exists when $%
\epsilon _{i}\neq 1/2$ (the Jacobian is nonzero).\quad For the general noise
model, the blue box in Fig. 1 is replaced by
\begin{equation}
Y^{y_{3}}X^{y_{2}}Z^{y_{1}}\propto X^{y_{3}+y_{2}}Z^{y_{3}+y_{1}}.
\end{equation}

\subsubsection{Anti-concentration condition}

The precise statement of anti-concentration condition is as follows \cite%
{Lund2017,Harrow2017,aaronson2011computational,PhysRevLett.117.080501,boixo2018characterizing,1706.03786}%
:
\begin{equation}
\underset{U\sim \mathcal{D}^{m}}{\mathbb{E}}\left[ \sum_{\mathbf{x}}p_{%
\mathbf{x}|U}^{2}\right] \leq \alpha 2^{-n},
\end{equation}%
where $\mathcal{D}$ denotes a uniform distribution for each blue box chosen
from the corresponding gate set, $m$ is the number of blue boxes, $\mathcal{D%
}^{m}$ denotes a joint distribution of $m$ random variables chosen
independently from $\mathcal{D}$, $p_{\mathbf{x}|U}$ is the probability
getting measurement result $\mathbf{x}$ for the circuit $U$, $n$ is the number of bits in the vector $\mathbf{x}$, and $\alpha$ is a constant. A sufficient
but not necessary condition to satisfy the anti-concentration condition is
that the ensemble achieves a unitary $2$-design which is believed to be
common for rich enough ensembles \cite%
{doi:10.1063/1.2716992,PhysRevA.80.012304,Brandao2016,1606.01914}.

\subsubsection{Bounds for truncating the Fourier series}

In the main text, we have given the approximation error of truncating
Fourier series through the average value and variance, using the Chebyshev
inequality to bound the error for most of the circuits in the ensemble.
Here, we give the detailed derivation. In this analysis, we assume the
anti-concentration condition to hold when we consider measurement on all the
output qubits.

For the variance, we have
\begin{eqnarray}
\Delta ^{2} &=&\mathbb{E}_{\mathbf{y}}[\delta _{\mathbf{y}}^{2}]-\delta
_{0}^{2}\leq \mathbb{E}_{\mathbf{y}}[\delta _{\mathbf{y}}^{2}]  \notag \\
&=&\frac{1}{2^{2m}}\sum_{\mathbf{y}}\left( \sum_{\mathbf{x}}|p_{\mathbf{x}|%
\mathbf{y}}^{\prime }-q_{\mathbf{x}|\mathbf{y}}^{\prime }|\right) ^{2}
\notag \\
&\leq &\frac{2^{r}}{2^{2m}}\sum_{\mathbf{x},\mathbf{y}}\left( p_{\mathbf{x}|%
\mathbf{y}}^{\prime }-q_{\mathbf{x}|\mathbf{y}}^{\prime }\right) ^{2}
\label{eq:bound} \\
&=&2^{2m+r}\sum_{\mathbf{x},\mathbf{y}}\left( p_{\mathbf{x},\mathbf{y}%
}^{\prime }-q_{\mathbf{x},\mathbf{y}}^{\prime }\right) ^{2}  \notag \\
&=&2^{2m+r}\sum_{\mathbf{x},\mathbf{s}}(\hat{p}_{\mathbf{x},\mathbf{s}%
}^{\prime }-\hat{q}_{\mathbf{x},\mathbf{s}}^{\prime })^{2}
\label{eq:parseval} \\
&=&2^{2m+r}\sum_{\mathbf{x},|\mathbf{s}_{1}|+|\mathbf{s}_{2}|\geq
l}(1-2\epsilon _{1})^{2|\mathbf{s}_{1}|}(1-2\epsilon _{2})^{2|\mathbf{s}%
_{2}|}\hat{q}_{\mathbf{x},\mathbf{s}_{1}\mathbf{s}_{2}}^{2}  \notag \\
&\leq &2^{2m+r}(1-2\epsilon )^{2l}\sum_{\mathbf{x},\mathbf{s}}\hat{q}_{%
\mathbf{x},\mathbf{s}}^{2}  \notag \\
&=&2^{2m+r}(1-2\epsilon )^{2l}\sum_{\mathbf{x},\mathbf{y}}\hat{q}_{\mathbf{x}%
,\mathbf{y}}^{2}  \notag \\
&=&(1-2\epsilon )^{2l}\frac{1}{2^{2m}}2^{r}\sum_{\mathbf{x},\mathbf{y}}q_{%
\mathbf{x}|\mathbf{y}}^{2}  \notag \\
&=&(1-2\epsilon )^{2l}\mathbb{E}_{\mathbf{y}}[2^{r}\sum_{\mathbf{x}}q_{%
\mathbf{x}|\mathbf{y}}^{2}]  \notag \\
&\leq &c^2(1-2\epsilon )^{2l}\leq c^2e^{-4\epsilon l}.  \notag
\end{eqnarray}%
If the number of measured qubits $r$ is a constant, since $\mathbb{E}_{\mathbf{y}}\left[
\sum_{\mathbf{x}}q_{\mathbf{x}|\mathbf{y}}^{2}\right] \leq 1$, we take $c^2=2^r$ which is also a constant. If we measure on all the $n$ qubits, i.e., $r=n$, we assume the
anti-concentration condition which means $\mathbb{E}_{\mathbf{y}}\left[
\sum_{\mathbf{x}}q_{\mathbf{x}|\mathbf{y}}^{2}\right] =\alpha 2^{-n}$, so $c^2=\alpha$, which is still a constant. The step of Eq. (%
\ref{eq:parseval}) holds as the Hadamard matrix is unitary which preserves the $%
l_{2}$ norm or equivalently we use the Parseval theorem in the theory of Fourier
analysis.

For the average value, let us consider its square:
\begin{eqnarray}
\delta^2_0&=&\left(\frac{1}{2^{2m}}\sum_{\mathbf{y}}\sum_{\mathbf{x}%
}|p^\prime_{\mathbf{x}|\mathbf{y}}-q^\prime_{\mathbf{x}|\mathbf{y}}|\right)^2
\notag \\
&\le&\frac{2^r}{2^{2m}}\sum_{\mathbf{x},\mathbf{y}}\left(p^\prime_{\mathbf{x}%
|\mathbf{y}}-q^\prime_{\mathbf{x}|\mathbf{y}}\right)^2,  \notag
\end{eqnarray}
which is the same as Eq. (\ref{eq:bound}). So, $\delta_0$ has the same bound
as $\Delta$, given by $ce^{-2\epsilon l}$.

According to the Chebyshev inequality, the outside fraction/probability $\eta$ is bounded by
\begin{eqnarray}  \label{eq:chebyshev}
\eta\equiv\Pr[\delta_{\mathbf{y}}>\delta]=\Pr[\delta_{\mathbf{y}%
}-\delta_0>\delta-\delta_0]\le\frac{\Delta^2}{(\delta-\delta_0)^2}.
\end{eqnarray}
By choosing
\begin{equation}  \label{eq:delta}
\delta=ce^{-\epsilon l}\gg \delta_0
\end{equation}
for sufficiently large $l$, we can bound $\eta$ by
\begin{equation}
\eta\le e^{-2\epsilon l}.
\end{equation}
This gives the statement below Eq. (7) in the main text and the result in theorem 1 when we choose $l=d$.










\textbf{Acknowledgements} We thank  Ignacio Cirac, Soonwon Choi, Liang Jiang, Zhengfeng Ji, Zhaohui Wei, Mingji Xia, Mengzhen Zhang, Sirui Lu and Yihong Zhang for helpful discussions. This work was supported by the Ministry of Education and the
National Key Research and Development Program of China (2016YFA0301902).

\textbf{Author contributions:}
X.G. carried out the work under L.M.D.'s supervision. All the authors made substantial contributions to this work.
The authors declare that they have no competing interests. Correspondence and requests for materials should
be addressed to L.M.D. (lmduan@tsinghua.edu.cn) or X. G. (gaoxungx@gmail.com).


\end{document}